\documentclass[aps,prb,showpacs,notitlepage,superscriptaddress,
floatfix,twocolumn]{revtex4-1}
\usepackage{bbm}
\usepackage{mathrsfs}
\usepackage{epsfig}
\usepackage{graphicx}
\usepackage{amsfonts}
\usepackage[figuresright]{rotating}
\usepackage{amssymb}
\usepackage{amsmath}
\usepackage{dcolumn}
\usepackage{bm}
\usepackage{braket}
\usepackage{comment}
\usepackage{mathtools}
\usepackage{xcolor}
\usepackage{multirow}
\usepackage{setspace}

\def\avg#1{\langle#1\rangle}

\def\be{\begin{equation}} \def\ee{\end{equation}}
\def\bea{\begin{eqnarray}} \def\eea{\end{eqnarray}}

\def\nn{\nonumber}

\usepackage{color}

\usepackage[normalem]{ulem}

\begin{document}
\title{High-order Time-Reversal Symmetry Breaking Normal State 
}
\author{Meng Zeng}
\affiliation{Department of Physics, University of California,
San Diego, California 92093, USA}
\author{Lun-Hui Hu}
\affiliation{Department of Physics, Zhejiang University, Hangzhou, China}
\author{Hong-Ye Hu}
\affiliation{Department of Physics, University of California,
San Diego, California 92093, USA}
\author{Yi-Zhuang You}
\affiliation{Department of Physics, University of California,
San Diego, California 92093, USA}
\author{Congjun Wu}
\email{wucongjun@westlake.edu.cn}
\affiliation{New Cornerstone Science Laboratory, Department of Physics, School of Science, Westlake University, 310024, Hangzhou, China}
\affiliation{Institute for Theoretical Sciences, WestLake University, 310024, Hangzhou, China}
\affiliation{Key Laboratory for Quantum Materials of Zhejiang Province, School of Science, Westlake University, 310024, Hangzhou, China}
\affiliation{Institute of Natural Sciences, Westlake Institute for Advanced Study, 310024, Hangzhou, China}

\begin{abstract}
Spontaneous time-reversal symmetry breaking plays an important
role in studying strongly correlated unconventional superconductors.
When two superconducting gap functions with different symmetries compete, the relative phase channel ($\theta_-\equiv \theta_1-\theta_2$) exhibits an Ising-type $Z_2$  symmetry due to the second order Josephson coupling,  where $\theta_{1,2}$ are the phases of two gap functions
respectively.
In contrast, the $U(1)$ symmetry in the
channel of
$\theta_+\equiv \frac{\theta_1+\theta_2}{2}$  is intact.
The phase locking, {\it i.e.}, ordering of $\theta_-$, can take place in the phase fluctuation regime before the onset of superconductivity, i.e. when $\theta_+$ is disordered.
If $\theta_-$ is pinned at $\pm\frac{\pi}{2}$, then 
time-reversal symmetry
is broken in the normal state, otherwise, if $\theta_-=0$, or, $\pi$, rotational
symmetry is broken, leading to a nematic normal state.
In both cases, the order parameters possess a 4-fermion structure beyond the scope of mean-field theory, which can be viewed as a high
order symmetry breaking. 
We employ an effective two-component $XY$-model assisted by a
renormalization group analysis to address this problem. As a natural by-product, we also find the other interesting intermediate phase corresponds to ordering of $\theta_+$ but with $\theta_-$ disordered.
This is the quartetting, or, charge-4e, superconductivity, which occurs above the low temperature $Z_2$-breaking charge-2e superconducting phase. 
Our results provide useful guidance for studying novel
symmetry breaking phases in strongly correlated superconductors.
\end{abstract}
\maketitle
\section{Introduction}
Unconventional superconductors (for instance, high-$T_c$ cuprates \cite{bednorz1986possible},
heavy-fermion systems \cite{stewart1984}, and
iron-based superconductors \cite{takahashi2008superconductivity})
have aroused a great deal of attentions for
novel symmetries in addition to the U(1) symmetry breaking.
Time-reversal symmetry (TRS) as well as parity and charge conjugation
are fundamental discrete symmetries, hence, spontaneous 
TRS-breaking superconductivity is of particular importance
\cite{lee2009,wu2010,stanev2010,khodas2012,garaud2014,
maiti2015,lin2016,wang2017topological,kang2018,wang2020,hu2020}.
Various TRS-breaking pairing structures are theoretically proposed,
including $d\pm id$ \cite{laughlin1998magnetic,tewari2008time},
$p\pm ip$ \cite{wang2018weak,wang2017topological},
$s\pm id$ \cite{lee2009,platt2012mechanism},
$p\pm is$ \cite{yang2020},
and $s+is$ \cite{silaev2017phase,hu2020},
and experimental evidence has been reported in various systems,
such as 
Re$_6$Zr \cite{singh2014,pang2018}, UPt$_3$ \cite{sauls1994,schemm2014},
PrOs$_4$Sb$_{12}$ \cite{aoki2003},
URu$_2$Si$_2$ \cite{mackenzie2003,schemm2015},
SrPtAs \cite{biswas2013},
LaNiC$_2$ \cite{hillier2009}, LaNiGa$_2$ \cite{hillier2012,weng2016},
Bi/Ni bilayers \cite{gong2017}, and CaPtAs \cite{shang2020}
(For details refer to a recent review \cite{ghosh2020a}.).
They are often probed by the zero-field $\mu$-spin relaxation, or, rotation \cite{schenck1985,lee1999,yaouanc2011}, and the polar
Kerr effect \cite{spielman1990,kapitulnik2009}.
TRS breaking signatures have also been reported in iron-based
superconductors \cite{grinenko2020,zaki2019}.

If TRS breaking arises from a complex pairing structure, it is often presumed that it develops after the onset of superconductivity.
However, these two transitions are of different nature: Superconductivity is of the $U(1)$ symmetry breaking and 
TRS is of $Z_2$, hence, they could take place at different temperatures.
It is interesting to further check whether TRS breaking
can occur before the superconducting transition as the temperature is lowered.
In fact, phase fluctuations are prominent
in strongly correlated superconductors above but close to $T_c$,
such as high $T_c$ cuprates \cite{emery1995importance} and
iron-based superconductors \cite{kasahara2016giant}.

In a two-gap superconductor,
the TRS breaking can be solely determined by
the relative phase between two gap functions.
The phases of two channels may fluctuate in a coordinated way
such that the relative phase is locked, leading to TRS 
breaking, while the total phase $\theta_+$ is disordered, hence, the system remains normal.
In the context of 2D bosons in the $p$-band, a TRS breaking Mott-insulating 
ground state was studied via the Ginzburg-Landau free energy analysis and the
quantum Monte Carlo simulatoins
\cite{cai2011,hebert2013}.
The TRS breaking normal state has been
studied in the context of three-gap
superconductors as a consequence from 
frustrations \cite{bojesen2013time}.

In this article, we show that there exists an
\textit{Ising symmetry breaking normal phase}
in a generic 2D two-gap superconductors when the gap functions
belong to different symmetries and are near degeneracy.
The key ingredient here, as mentioned above, is the superconduct phase fluctuations.
Hence, it is a phase-fluctuation induced
TRS-breaking, or, a nematic normal state.
By the symmetry principle, the two gap functions couple via a second
order Josephson term. Therefore, we dub the resultant symmetry-breaking normal state as the ``high-order'' symmetry-breaking state.
In the phase fluctuation regime, the low energy physics is described by
a coupled two-component $XY$-model, which is mapped to a
coupled sine-Gordon model and analyzed by the renormalization group
(RG) method.
Unlike the small difference in the superconducting transition temperature and
the TRS-breaking temperature obtained in Ref.[\onlinecite{bojesen2013time}]
from the frustration effects in the three-band model,
the phase-locking, or, the $Z_2$ symmetry breaking temperature can be considerably
larger than the superconducting $T_c$.
Another competing order, the quartetting \cite{wu2005competing}, or, charge-4e phase
\cite{berg2009charge}, can also appear above $T_c$, which corresponds to ordered total phase $\theta_+$ but with the relative phase $\theta_-$ disordered, i.e. the $U(1)$ symmetry in the $\theta_+$ channel is broken whereas the $Z_2$ symmetry in the $\theta_-$ channel is preserved. 
All these phases exhibit the 4-fermion-type order parameters,
and thus are difficult to analyze in mean-field theories.
Quite remarkably, the $Z_2$-breaking TRS-breaking normal state has recently been experimentally observed in hole-doped $\text{Ba}_{1-x}\text{K}_x\text{Fe}_2\text{As}_2$ \cite{vadim2021state,vadim2022calorimetric}, where the TRS-breaking transition is identified with the onset of specific-heat anomaly and spontaneous Nernst signal is also detected in the TRS-breaking normal state. The $Z_2$-breaking nematic normal state has been observed in $\text{Sr}_2\text{RuO}_4$ \cite{harter2023electronic} using optical anisotropy measurement. Even though the normal state nematicity most likely has a different origin from our theory because it can happen at much higher temperature scale, the same experimental techniques can be used to detect nematicity in the phase fluctuation regime proposed in our work.
The competing charge-4e state has also been observed recently in kagome superconductor $\text{CsV}_3\text{Sb}_5$ \cite{charge4e2022discovery}, where the quantization of magnetic flux in units of $hc/4e$ is observed.

The paper is structured as the following: In Sec.~\ref{sec:GL-theory} we introduce the Ginzberg-Landau theory for superconductors with two gap functions of different symmetries. 
They couple due to the second order Josephson effect.
In Sec. \ref{sect:fluctuations},
we focus on the phase degree of freedom by mapping the theory to a coupled $XY$-model, which can be further mapped to a coupled sine-Gordon model, setting the stage for the RG study. 
In Sec.~\ref{sec:RG-analysis}, we perform detailed RG analysis of the sine-Gordon model by considering the effects of various symmetry-allowed couplings between different channels, which lead to the emergence of different phase diagram topologies. In Sec.~\ref{sec:application}, we briefly discuss the application of our theory to Fe-based superconductors. Then we conclude in Sec.~\ref{sec:conclusion}.

\section{Ginzberg-Landau theory 
with two gap functions}
\label{sec:GL-theory}
We start with the Ginzberg-Landau (GL) free-energy of superconductivity
with two gap functions.
Each one by itself is time-reversal invariant.
These two gap functions belong to two different representations
of the symmetry group, say, the $s$-wave and $d$-wave symmetries of a tetragonal system, or, different components of a two-dimensional representation, say, the $p_x$ and $p_y$-symmetries.
They cannot couple at the quadratic level since
no invariants can mix them at this level.
Bearing this in mind, the GL free-energy is constructed as
$\mathcal{F}=\mathcal{F}_1+\mathcal{F}_2$ with
\bea
\mathcal{F}_1&=&\gamma_1 |\vec \nabla \Delta_1|^2
+ \gamma_2 |\vec \nabla \Delta_2|^2+
\alpha_1 (T) |\Delta_1|^2+\alpha_2 (T)|\Delta_2|^2 \nn \\
&+&\beta_1|\Delta_1|^4
+\beta_2|\Delta_2|^4
+\kappa |\Delta_1|^2|\Delta_2|^2,
\label{Eq:free energy 1}
\\
\mathcal{F}_2&=& \lambda\left(\Delta_1^2\Delta_2^{*2}
+\Delta_1^{*2}\Delta_2^2\right),
\label{Eq:free energy 2}
\eea
where $\alpha_{1,2}(T)$ are functions of temperatures,
and their zeros determine their superconducting transition temperatures
when the two gap functions decouple.
$\gamma_{1,2}$, $\beta_{1,2}$ are all positive and $\kappa^2<4\beta_1\beta_2$
to maintain the thermodynamic stability.
If the gap functions form a two-dimensional representation
of the symmetry group,
then $\alpha_{1}=\alpha_2$, $\beta_1=\beta_2$, and
$\gamma_1=\gamma_2$, otherwise, they are generally independent.
Nevertheless,  we
consider the case that they are nearly degenerate, i.e.,
$\alpha_1\approx \alpha_2$, when they belong to two
different representations, such that they can coexist.

The $\mathcal{F}_1$-term only depends on the magnitude of $\Delta_{1,2}$,
hence, is phase insensitive.
We assume that the two gap functions can form a quartic invariant
as the $\mathcal{F}_2$-term, as in the cases of $s$ and $d$-waves,
and $p_x$ and $p_y$-waves.
The $\mathcal{F}_2$-term does depend on the relative phase between $\Delta_{1,2}$,
which can be viewed as a 2nd order Josephson coupling.
To minimize the free energy, the relative phase between two gap
functions  $\theta_-=\theta_1-\theta_2 =\pm \frac{\pi}{2}$ at $\lambda>0$, i.e.,
they form $\Delta_1\pm i \Delta_2$, breaking TRS
spontaneously.
On the other hand, when $\lambda<0$, $\theta_-=0$, or, $\pi$.
They form the nematic superconductivity $\Delta_1\pm \Delta_2$,
breaking the rotational symmetry.
The magnitude of the mixed gap function remains isotropic in momentum
space in the former case, while that in the latter case is anisotropic.
The value of $\lambda$ depends on the energetic details of a
concrete system.
At the mean-field level, the free energy is a convex functional of the
gap function distribution in the absence of spin-orbit coupling
\cite{cheng2010,wu2010,yang2020}.
This favors a relatively uniform distribution of gap function
in momentum space, corresponding to the complex mixing
$\Delta_1\pm i\Delta_2$ , i.e., $\lambda>0$.
Nevertheless, the possibility of $\lambda<0$ cannot be ruled out,
which could take place in the presence of spin-orbit coupling
\cite{wang2017topological}, or as a result beyond the mean-field BCS theory.
This leads to the gap function $\Delta_1\pm \Delta_2$, which
breaks the rotational symmetry leading to nematic
superconductivity.

\section{New phases due to the
phase fluctuations}
\label{sect:fluctuations}

The above GL analysis only works in the superconducting phases
in which both $\Delta_{1,2}$ develop non-zero expectation values.
However, it does not apply to the phase fluctuation regime
above $T_c$.
Let us parameterize the gap functions as $\Delta_{1,2}=|\Delta_{1,2}|e^{i\theta_{1,2}}$.
In the phase fluctuation regime, the order magnitudes $|\Delta_{1,2}|$ are already
significant, and their fluctuations can be neglected.
On the contrary, the soft phase fluctuations dominate the
low energy physics, and the system remains in the normal
state before the onset of the long-range phase coherence.

New states can arise in the phase fluctuation regime in which
neither of $\Delta_{1,2}$ is ordered.
A possibility is that the system remains in the normal state
but $\theta_-$ is pinned:
If $\theta_-= \pm\frac{\pi}{2}$, then $\mbox{Im}\Delta_1^*\Delta_2$
is ordered, which breaks TRS; if $\theta_-= 0,\pi$, then $\mbox{Re}\Delta_1^*\Delta_2$ is ordered, which breaks rotation
symmetry.
Similar physics occurs in the $p$-orbital band Bose-Hubbard model, where the boson operators in the $p_{x,y}$-bands play the role
of $\Delta_{1,2}$, respectively.
The transitions of superfluidity and TRS breaking divide the phase diagram
into four phases of superfluidity states with and without TRS breaking,
and the Mott insulating state with and without TRS breaking,
where TRS here corresponds to the development of the onsite
orbital angular momentum by occupying the complex orbitals
$p_x\pm ip_y$
\cite{wu2009unconventional,hebert2013exotic}.
The TRS-breaking normal states were also studied in the context of
competing orders in superconductors \cite{fernandes2019intertwined, fischer2016fluctuation}.
Another possibility is that the total phase $\theta_+=\theta_1+\theta_2$
is pinned, i.e., $\Delta_1\Delta_2$ is ordered.
This corresponds to the quartetting instability ,
i.e., a four-fermion clustering instability analogous to the
$\alpha$-particle in nuclear physics.
The competition between the pairing and quartetting instabilities
in one dimension has been investigated by one of the authors \cite{wu2005competing}.
Later it was also studied in the context of high-T$_c$ cuprates
as the charge-4e superconductivity \cite{berg2009charge}.

However, all the above states involve order parameters consisting
of 4-fermion operators.
Hence, they are beyond the ordinary mean-field theory based on
fermion bilinear order parameters.
To address these novel states, we map the above GL free-energy
to the $XY$-model on a bilayer lattice, and perform the renormalization
group (RG) analysis to study the possible phases.
Since there should be no true long-range order of the U(1) symmetry
at finite temperatures, we mean the quasi-long-ranged ordering
of the Kosterlitz--Thouless (KT) transition.
The model is expressed as
\bea
H&=&-J_1\sum_{\langle i,j \rangle}\mathrm{cos}(\theta_{1i}-\theta_{1j})-J_2\sum_{\langle i,j \rangle}\mathrm{cos}(\theta_{2i}-\theta_{2j})\nn \\
&+&\lambda^\prime\sum_i\mathrm{cos}2(\theta_{1i}-\theta_{2i}),
\label{Eq:hamiltonian}
\eea
where $\theta_{1,2}$ are compact U(1) phases
with the modulus $2\pi$.
$J_{1,2}$ are the intra-layer couplings estimated as
$J_{1,2}\approx \gamma_{1,2}|\Delta_{1,2}|^2$,
and $\lambda^\prime$ is the inter-layer coupling estimated as
$\lambda^\prime\approx 2\lambda|\Delta_1|^2|\Delta_2|^2$.

Following the dual representation of the 2D classic $XY$-model as 
detailed in the Appendix~\ref{append:duality}, the
above model Eq. (\ref{Eq:hamiltonian}) can be mapped to the following
multi-component sine-Gordon model, which is often employed for
studying coupled Luttinger liquids \cite{wu2003competing,hu2019interacting}.
Its Euclidean Lagrangian in the continuum is defined as
$L=\int d^2x \mathcal{L}(x)$ \cite{fradkin2013field}, where
\bea
\mathcal{L}(x)&=&\frac{1}{2K_1}\left(\partial_{\mu}\phi_1 \right)^2+\frac{1}{2K_2}
(\partial_{\mu}\phi_2)^2  +g_{\theta_-}\mathrm{cos}2\left(\theta_1-\theta_2\right)\nn \\
 &-&g_{\phi_1} \mathrm{cos}2\pi\phi_1
 -g_{\phi_2} \mathrm{cos}2\pi\phi_2,
 \label{Eq:lagrangian1}
\eea
where $\phi_{1,2}$ are the dual fields to the superconducting phase fields of $\theta_{1,2}$ with commutation relations $[\theta_{1,2}(t,x),\partial_y\phi_{1,2}(t,y)]=2\pi i \delta(x-y)$, and the Luttinger parameters $K_{1,2}=J_{1,2}/T$.
(Please note that $K_{1,2}$ appear in the 
denominators in Eq. \ref{Eq:lagrangian1}
since we are using the dual representation.)
The compact radius of $\theta_{1,2}$ is $2\pi$, and that of the
vortex fields $\phi_{1,2}$ is 1.
$g_{\theta_-}$ is proportional to $\lambda^\prime$
in Eq. (\ref{Eq:hamiltonian});
$g_{\phi_1,\phi_2}$ are proportional to the vortex fugacities of the
phase fields $\theta_{1,2}$, respectively.
For simplicity, all of these $g$-eology coupling constants
have absorbed the short-distance cutoff of the lattice.

\section{Renormalization group analysis for phase diagrams}
\label{sec:RG-analysis}
In this section, we explore the possible phase diagrams using RG analysis for the case where the two channels are degenerate,
i.e., $J_1=J_2\equiv J$, $g_{\phi_1}=g_{\phi_2}\equiv \frac{1}{2}g_{\phi_\pm}$ and $K_1=K_2=J/T\equiv K$.

Due to the permutation symmetry between these two channels, the coupled theory is rewritten in terms of the collective basis $\theta_{\pm}$, $\phi_{\pm}$ channels conveniently defined as 
\bea
\theta_{+}&\equiv& (\theta_1+\theta_2)/2, 
~\theta_{-}\equiv \theta_1-\theta_2 \nn \\
\phi_{+} &\equiv& \phi_1+\phi_2, \ \ \
\ \ \ \phi_-\equiv (\phi_1-\phi_2)/2.
\label{eq:evenodd}
\eea
The compact radius of $\theta_{\pm}$ can be chosen as 
$2\pi$, and that of the vortex fields $\phi_{\pm}$
remains 1.
This new basis is also convenient in the sense that it makes the symmetries of the coupled system explicit and at the same time it preserves the commutation relations between the fields and the dual fields, i.e. $[\theta_{\pm}(t,x),\partial_y\phi_{\pm}(t,y)]=2\pi i \delta(x-y)$. 
The Lagrangian has a $U(1)$ symmetry in $\theta_+$ channel, $\theta_+\to\theta_++\alpha$ with $\alpha\in[0,2\pi)$, and the $Z_2$ symmetry in the $\theta_-$ channel, $\theta_-\to\theta_-+\pi$, due to the $\cos2\theta_-$ term. 

Based on symmetry alone, there can exit four different phases: (i) Both $U(1)$ and $Z_2$ are unbroken, i.e. the normal phase; (ii) Only  $Z_2$ is broken, i.e. TRS-breaking (or nematic) normal phase; (iii) Only $U(1)$ is broken, i.e. the charge-$4e$ phase; (iv) Both $U(1)$ and $Z_2$ are broken, i.e. the TRS-breaking (or nematic) superconducting phase.
We can start with the free theory containing only the kinetic terms, and then add on the most relevant symmetry-preserving interaction terms to obtain phase diagrams containing all of the four possible phases discussed above, but with different phase diagram topologies. 

With the basis transformation defined above, the free part of the Lagrangian in Eq. (\ref{Eq:lagrangian1}) can be equivalently written in the $\theta_\pm,\phi_\pm$ basis as,
\begin{equation}
\mathcal{L}_0(x)=\frac{1}{4K_+}(\partial_{\mu}\phi_+)^2+\frac{1}{K_-}
(\partial_{\mu}\phi_-)^2,
\label{Eq:lagrangian2}
\end{equation}
where the initial values of both $K_\pm$ are both $J/T$.

Once various interaction terms are added, the phase diagram lives in a high dimensional parameter space. As a result, it is difficult to present a complete phase diagram involving all the parameters. 
However, based on the symmetry analysis provided above, there are only four phases in total. 
Therefore, it is possible to show two dimensional (2D) slices of the phase diagram that contains the four phases. 
Interestingly enough, topologically distinct configurations of phase boundaries can be obtained, depending on which interaction terms dominate at low energy. 
In the following two subsections, we will present two generic cases showing three types of phase diagram topologies.

\subsection{$\phi_\pm$ channels decoupled}
We consider possible local vortex terms in the collective basis, which are discussed in Appendix \ref{append::Kmatrix}. 
The most relevant one is  $g_{\phi_{int}}\cos\pi\phi_+\cos2\pi\phi_-$, which couples the even and odd channels together.
It originates from the vortex fugacity terms in the individual basis $\cos2\pi\phi_1+ \cos 2\pi \phi_2$.
The sign change of $\cos\pi\phi_+\cos2\pi\phi_-$ from shifting $\phi_+$ by 1 can be compensated by a shift of $\phi_-$ by $1/2$, and vice versa.
The next leading vortex terms are $g_{\phi_+}\cos2\pi\phi_+$ and $g_{\phi_-}\cos4\pi\phi_-$ in the even and odd channels respectively, which originate from the inter-layer vortex-vortex coupling in the original basis $\cos2\pi\phi_1\cos2\pi\phi_2\pm\sin2\pi\phi_1\sin2\pi\phi_2$.

We begin with the limit that the initial value of the interlayer 
phase coupling $g_{\theta_-}$ is large.
In this case, vortices in two layers tend to be aligned together. 
Hence, the independent single vortex excitation in each 
layer is not favored and its fugacity is suppressed, 
{\it i.e.}, $|g_{\phi_\pm}|\gg |g_{\phi_{int}}|$.
In this limit, the $g_{\phi_{int}}$-term is neglected,
then the system is decoupled in the collective 
basis with Lagrangian given by,
\begin{equation}
\begin{split}
\mathcal{L}_1(x)&=\frac{1}{4K_+}(\partial_{\mu}\phi_+)^2+\frac{1}{K_-}
(\partial_{\mu}\phi_-)^2 \\
&-g_{\phi_+} \mathrm{cos}2\pi\phi_+
-g_{\phi_-} \mathrm{cos}4\pi\phi_-+g_{\theta_-}\mathrm{cos} 2\theta_-.
\end{split}
\end{equation}

In this decoupled case, we expect the $U(1)$-breaking transition in the $\theta_+$ channel to be completely independent from the $Z_2$-breaking transition in the $\theta_-$ channel. The phase diagram can be obtained by numerically solving the following set of RG equations (see Appendix~\ref{append:RG-derivation} for details), 
\begin{equation}
\begin{split}
&\frac{\mathrm{d}g_{\phi_+}}{\mathrm{d}\ln l}
=\left(2-2\pi K_+ \right)g_{\phi_+},\\
&\frac{\mathrm{d}g_{\phi_-}}{\mathrm{d}\ln l}
=\left(2-2\pi K_- \right)g_{\phi_-},\\ 
&\frac{\mathrm{d}g_{\theta_-}}{\mathrm{d}\ln l}=\left(2-\frac{2}{\pi K_-}
\right)g_{\theta_-},\\
&\frac{\mathrm{d}K_+}{\mathrm{d}\ln l}
=-2\pi^3g_{\phi_+}^2K_+^2,\\
&\frac{\mathrm{d}K_-}{\mathrm{d}\ln l}
=-4\pi^3g_{\phi_-}^2K_-^2+4\pi g_{\theta_-}^2,
\label{Eq:RG_L1}
\end{split}
\end{equation}
where both of the initial values of $K_\pm$ are $J/T$. 

Below we analyze the nature of the fixed points of RG
for four different phases: (I) the $Z_2$ breaking SC 
phase; (II) $Z_2$ breaking normal phase; (III) quartetting
phase; (IV) normal phase.
The values of couplings at these fixed points are
summarized in Table. \ref{tab1}.

Phase I and Phase II are the $Z_2$ breaking superconducting (SC) and normal phases, respectively.
In the former case,
the relative phase $\theta_-$ is locked, while the $\theta_+$ is quasi-long-range ordered.
Hence, $g_{\theta_-}\to \infty$ and correspondingly $K_-\to \infty$.
As for the vortex term $\cos 2\pi \phi_+$ in the $\phi_+$ channel, 
such a vortex term should be irrelevant in phase I, 
which requires $K_+$ takes a constant value
with $K_+>1/\pi$, and $(g_{\phi_-},g_{\phi_+}) \to 
(0,0)$.
In the $Z_2$-breaking normal phase,
the relative phase $\theta_-$ remains locked, while
the vortex $\phi_+$ proliferates such that
superfluidity is lost.
Notice that the $Z_2$-breaking normal 
state appears in the intermediate temperature, i.e. the phase fluctuations of the underlying SC state  lead to the symmetry-breaking normal state above the SC critical temperature. 
This intermediate phase can be the TRS breaking state, or, the nematic state depending on the $\theta_-$ is pinned at $\pm \frac{\pi}{2}$, or, 
$0$ or $\pi$, respectively.
In such a phase, $g_{\theta_-}\to \infty$ and
$K_-\to +\infty$, which are the same as in phase I.
On the other hand, in order to proliferate vortices
in the $\phi_+$ channel, $g_{\phi_+} \to \infty$, which
means $K_+ \to 0$.
Then $(g_{\phi_-},g_{\phi_+}) \to (0,\infty)$.

Phase III and Phase IV, {\it i.e.}, the quartetting ($4e$) state and the normal state respectively, are
both the $Z_2$-symmetric phases. 
For the quartetting ($4e$) state,
the vortex field in the relative channel $\phi_-$ 
condenses, while the $\theta_+$ channel is 
quasi-long-range ordered.
The condensation of $\phi_-$ means that $g_{\theta-} \to 0$ and $K_-\to 0$, and $g_{\phi-} \to \infty$.
The quasi-long-range ordering of $\theta_+$ requires
$g_{\phi_+}\to 0$, which means that the renormalized value 
of $K_+$ reaches a constant with $K_+>\frac{1}{\pi}$, which
becomes a line of stable fixed points.
As for the normal state, it means that 
the vortex fields in both channels condense.
This simply gives rise to 
$g_{\phi_+}\to \infty$,
$g_{\phi_-}\to \infty$, and $g_{\theta-} \to 0$,
which corresponds to $K_+\to 0$ and $K_-\to 0$.

\begin{figure}[h]
\centering
\includegraphics[width=0.4\textwidth]{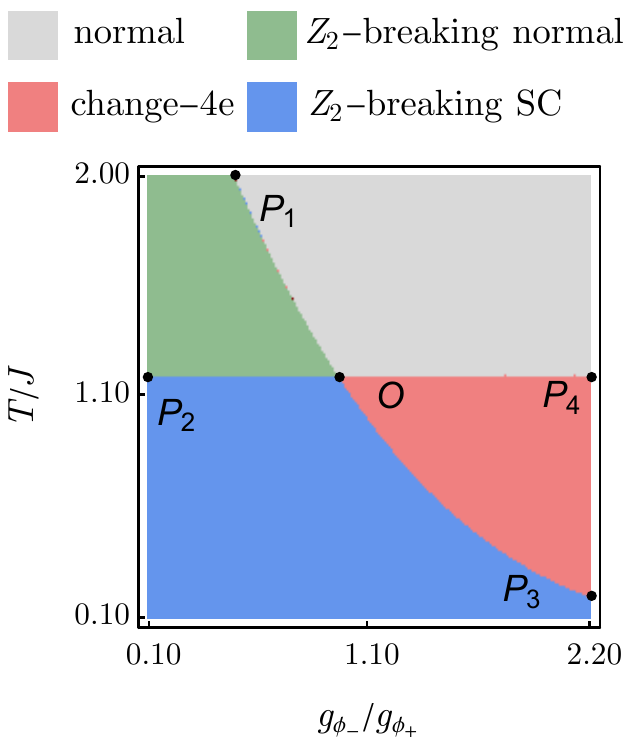}
\caption{Phase diagram {\it v.s.} temperature and $g_{\phi_-}/g_{\phi_+}$ by numerically
integrating the RG Eq. (\ref{Eq:RG_L1}).
The initial values of coupling constants are $g_{\phi_+}=0.2$ and $g_{\theta_-}=0.01$. All of the four phases appear and meet at the multi-cirtical point $O$.}
\label{fig:phase1}
\end{figure}

By numerically integrating the RG Eq.(\ref{Eq:RG_L1}), the above
four phases are obtained.
The phase diagram as a function of temperature and fugacity ratio between two channels $g_{\phi_-}/g_{\phi_+}$ is shown in Fig.~\ref{fig:phase1}. 
The fixed point values of the couplings deep in the four phases as well as on the phase boundaries are listed in Table~\ref{tab1}. 
As expected, when the two channels are decoupled, the $U(1)$-breaking phase boundary and the $Z_2$-breaking phase 
boundary are independent from each other and crosses at a single point, diving the phase diagram into four regions characterized by different symmetry breaking patterns.

\begin{table}[h]
    \centering
    \begin{tabular}{l|c c c c c c c c}
    \hline
       phases and\\ phase boundaries 
       & $g_{\theta_-}$ &$g_{\phi_-}$& $g_{\phi_+}$ & $K_-$&$K_+$ \\
          \hline
    \hline
(I)   $Z_2$-breaking SC &$\infty$& 0 & 0 & $+\infty$ &$>\frac{1}{\pi}$\\  
(II)  $Z_2$-breaking normal & $\infty$ & 0 & $\infty$ &$+\infty$ &0 \\ 
 (III) quartetting (4$e$) &$0$ & $\infty$&0 &0& $>\frac{1}{\pi}$\\ 
(IV)  normal & 0 & $\infty$ &$\infty$ &0 &0\\   
      \hline
    \hline
    $P_1O$ &  \multicolumn{2}{c}{$\frac{g_{\phi_-}}{g_{\theta_-}}=\pm 1$}
    &$\infty$ &$\frac{1}{\pi}$& 0\\
    $P_2O$&$\infty$  &0&0  & $+\infty$& $\frac{1}{\pi}$\\
    $P_3O$&\multicolumn{2}{c}{$\frac{g_{\phi_-}}{g_{\theta_-}}=\pm 1$} &
    0& $\frac{1}{\pi}$ &$\frac{1}{\pi}$\\
    $P_4O$&0&$\infty$& 0 &0&$\frac{1}{\pi}$\\
     \hline 
    \end{tabular}
\caption{Values of couplings at fixed points in the four phases and on phase boundaries under RG Eq.(\ref{Eq:RG_L1}).
}
\label{tab1}
\end{table}


Along the phase boundary $P_1 P_3$ (excluding the multi-critical point $O$), it represents a $Z_2$-breaking transition inside the normal state with $K_-=\frac{1}{\pi}$. 
Then the fixed point condition for $K_-$ can be solved to give the relation $|g_{\phi_-}|=|g_{\theta_-}|$. 
The segment of $P_1O$ lies in the
normal state with 
$g_{\phi_+}=\infty$ with $K_+\to 0$ separating the $Z_2$-breaking
normal state and the complete normal state.
In contrast, the $P_3O$ lies in the region with quasi-long-range ordered U(1) phase $\theta_+$ separating the $Z_2$-breaking SC state with $\theta_-$ locked and the quartetting charge-$4e$ phase. 
The boundary of $P_2P_4$ separates the superfluid phase and the normal phase, below which the U(1) phase $\theta_+$ becomes quasi-long-range ordered.
The line of $P_2O$ marks the boundary between the $Z_2$-breaking normal and SC phases.
Similarly, the line of $P_4O$ marks the boundary between the quartetting phase and the normal phase. 

Here we comment on the exact duality on the critical line $P_1P_3$. More precisely, it is the duality between the field $\theta_-$ and its dual $\phi_-$.
To make the duality manifest, we can do a field rescaling 
\bea
\tilde{\phi}_-\equiv \sqrt{2\pi}\phi_-, 
\ \ \
\tilde{\theta}_-\equiv \theta_-/\sqrt{2\pi},
\eea
such that the two mass terms become $\cos\sqrt{8\pi}\tilde{\phi}_-$ and $\cos\sqrt{8\pi}\tilde{\theta}_-$ respectively. At the same time, the Luttinger parameter also has to be rescaled $\tilde{K}_-\equiv \pi K_-$, which becomes 1 at the critical point. It is again straightforward to show that  $|g_{\phi_-}|=|g_{\theta_-}|$ at criticality. Then the duality of exchanging $\tilde{\theta}_-$ and $\tilde{\phi}_-$  on the Lagrangian level is made explicit. Such a theory has also been studied as the field theory description of one dimensional deconfined quantum critical point with $Z_2\times Z_2$ symmetry\cite{zhang2023exactly}. In our case, the first $Z_2$ acts on the field $\theta_-$, and the second $Z_2$ acts on its dual $\phi_-$. The mixed anomaly between the two $Z_2$ symmetries dictates that when one is preserved the other has to be spontaneously broken. It has also been shown that such exotic critical point can be mapped to the usual Landau symmetry-breaking transition of a 1d $Z_4$ clock model, whose critical point is just two decoupled copies of Ising CFT\cite{zhang2023exactly,su2023boundary}. It is interesting to note that such exotic critical point can arise naturally in the two-gap superconductors that we study.

\subsection{$\phi_\pm$  coupled through $\cos\pi\phi_+\cos2\pi\phi_-$}

Now we add the vortex term of $\cos\pi\phi_+\cos2\pi\phi_-$ which couples the $\phi_\pm$ fields together.
The following Lagrangian is obtained,
\begin{equation}
\begin{split}
\mathcal{L}_2(x)&=
\frac{1}{K_-} (\partial_{\mu}\phi_-)^2 +
\frac{1}{4K_+}(\partial_{\mu}\phi_+)^2 \\
&+g_{\theta_-}\mathrm{cos} 2\theta_-
-g_{\phi_{int}} \mathrm{cos}\pi\phi_+ 
\mathrm{cos}2\pi\phi_- \\
&
-g_{\phi_-} \mathrm{cos}4\pi\phi_-
-g_{\phi_+} \mathrm{cos}2\pi\phi_+.
\end{split}
\end{equation}

The RG equations can be written down as the following (see Appendix~\ref{append:RG-derivation}),
\begin{eqnarray}
\begin{split}
&\frac{\mathrm{d}g_{\theta_-}}{\mathrm{d}\ln l}=\left(2-\frac{2}{\pi K_-}
\right)g_{\theta_-},\\
&\frac{\mathrm{d}g_{\phi_{int}}}{\mathrm{d}\ln l}=\left[2-\frac{\pi}{2} 
\left(K_+ + K_-\right)\right] g_{\phi_{int}}, \\ \\
&\frac{\mathrm{d}K_-}{\mathrm{d}\ln l}
=-4\pi^3K_-^2\left(g_{\phi_-}^2+g_{\phi_{int}}^2/8\right)+4\pi g_{\theta_-}^2 \\ 
&\frac{\mathrm{d}K_+}{\mathrm{d}\ln l}
=-4\pi^3K_+^2\left(g_{\phi_+}^2+g_{\phi_{int}}^2/8\right),\\ \\
&\frac{\mathrm{d}g_{\phi_-}}{\mathrm{d}\ln l}
=\left(2-2\pi K_- \right)g_{\phi_-} +\frac{\pi}{4}g^2_{\phi_{int}},\\
&\frac{\mathrm{d}g_{\phi_+}}{\mathrm{d}\ln l}
=\left(2-2\pi K_+ \right)g_{\phi_+} +\frac{\pi}{4}g^2_{\phi_{int}},\\
\label{Eq:RG_L2}
\end{split}
\end{eqnarray}

By analyzing Eq. (\ref{Eq:RG_L2}), again we have the four stable phases as discussed in the decoupled case in the previous section before. 
The values of couplings at the fixed points corresponding to
these phases and at the phase boundaries are summarized in Table. \ref{Tab_2}.
Compared to the decoupled case, the $Z_2$-breaking SC phase,  
the $Z_2$-breaking normal phase, and the quartetting phase 
further require that $g_{\phi_{int}} \to 0$.
Furthermore, the quartetting phase requires $K_+>\frac{4}{\pi}$
to ensure the irrelevancy of the $g_{\phi_{int}}$-term.
As for the normal phase, certainly $g_{\phi_{int}} \to \infty$.

\begin{table}[h]
\centering
\begin{tabular}{l |c llc c c }
\hline
Phases  & $g_{\theta_-}$ &$g_{\phi_-}$ &$g_{\phi+}$ &$g_{\phi_{int}}$ &$K_+$&$K_-$ \\
       \hline
       \hline
(I) $Z_2$-breaking SC& $\infty$  &0 &0&0&$>\frac{1}{\pi}$&$+\infty$\\
(II) $Z_2$-breaking normal& $\infty$ &0  &$\infty$&0 &0&$+\infty$
\\      
(III) Quartetting (4$e$) & $0$  &$\infty$  &0 &0& $>\frac{4}{\pi}$&0
\\   
(IV) Normal  & 0 &$\infty$  &$\infty$ &$\infty$ &0&0 \\    
     \hline
     \hline
          $P_1O_1$ & $\infty$  &$\infty$ &$\infty$& $\infty$& 0&$\frac{1}{\pi}$\\
          $P_2O_1$ &$\infty$ &0 &0& 0& $\frac{1}{\pi}$&0\\  
      $P_3O_2$ & \multicolumn{2}{c}{$\frac{g_{\phi_-}}{g_{\theta_-}}=\pm 1$}  &0& 0& $>\frac{4}{\pi}$&$\frac{1}{\pi}$\\
       $P_4O_2$ &0 &$\infty$ &0&0& $\frac{1}{\pi}$&0\\
      $O_1O_2$ & \multicolumn{2}{c}{$\frac{g_{\phi_-}}{g_{\theta_-}}=\pm 1$}  &0& 0& $\frac{1}{\pi}$&$\frac{1}{\pi}$\\
     \hline
    \end{tabular}
\caption{The values of couplings at the fixed
points corresponding to four stable phases and on the phase boundaries by solving Eq.~(\ref{Eq:RG_L2}). 
}
\label{Tab_2}
\end{table}

A key feature of the new phase diagram after introducing the $g_{\phi_{int}}$ term is that the previous tetra-critical point $O$ splits into a pair of tri-critical points $O_1$ and $O_2$, such that there appears a direct transition across $O_1O_2$ from the $Z_2$-breaking 
SC phase to the normal state \cite{gmzhang2022phase}. 

A small $g_{\phi_{int}}$-term does not change the boundaries much when deep inside the $Z_2$-ordered or the superconducting regions as long as they are relatively far away from $O_1O_2$.
In this case, the RG processes in the two channels can be decomposed into fast and slow steps.
For example, along the boundary $P_2O_1$ deep inside the $Z_2$-breaking phase, $\theta_-$ is pinned, which renders the $g_{\phi_{int}}$-term highly irrelevant by disordering the $\phi_-$ field. 
Similarly, along the boundary $P_3O_2$ deep inside the superfluid phase, $g_{\phi_+}$ is quickly suppressed to 0. 
The RG process in the $\phi_+$ channel stops quickly, such that $g_{\phi_{int}}$ does not grow much and remains small still.
Furthermore, $\phi_+$ remains power-law fluctuating, 
which suppresses the effect of the $g_{\phi_{int}}$-term. 

\begin{figure}[h]
\centering
\includegraphics[width=0.4\textwidth]{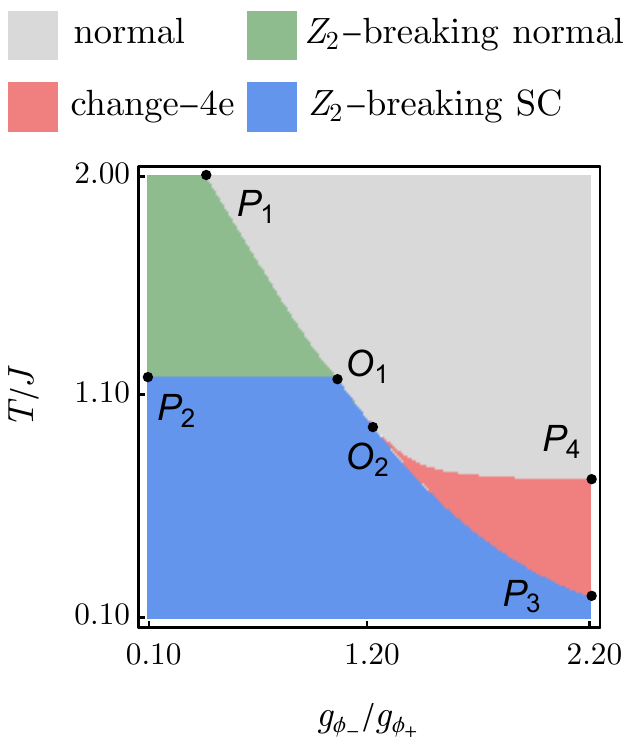}
\caption{Phase diagram {\it v.s.} temperature and $g_{\phi_-}/g_{\phi_+}$ by numerically
integrating the RG Eq. (\ref{Eq:RG_L2}).
The initial values of coupling constants are $g_{\phi_+}=0.2,g_{\theta_-}=0.01$ and 
$g_{\phi_{int}}=0.001$. 
Different from Fig.~\ref{fig:phase1}, this phase diagram features a direct transition boundary $O_1O_2$ between the normal state and the $Z_2$-breaking SC phase due to the coupling between the $\phi_\pm$ channels.
}
\label{fig:phase2}
\end{figure}

On the other hand, the $g_{\phi_{int}}$-term affects the boundaries surrounding the normal phase. 
As for the part along $P_4O_1$ deeply inside the $Z_2$-disordered
region, $\phi_-$ is pinned.
The $g_{\phi_{int}}$-term becomes $g^\prime \cos \pi \phi_+$, which is a half-quantum vortex with a renormalized coupling constant
$g^\prime=g_{\phi_{int}}\avg{\cos2\pi\phi_-}$.
Such a term is more relevant than the one-vortex term of $g_{\phi+}$ although its coupling is weaker. 
Nevertheless, it extends the region of the normal state significantly
as shown in Fig. \ref{fig:phase2}.
As for $P_1O_1$ deep inside the normal phase, $g_{\phi_+}$-term reaches the order of 1 quickly, and $\phi_+$ is pinned. 
Then the $g_{\phi_{int}}$-term becomes $g^{\prime\prime} 
\cos 2\pi \phi_-$ with $g^{\prime\prime}=g_{\phi_{int}}\avg{\cos\pi\phi_+}$, which is more relevant than the existing $g_{\phi_-}\cos4\pi\phi_-$ term. 
It changes the competition between the condensation of $\theta_-$ and $\phi_-$, which corresponds to the $Z_2$-ordered and disordered state, respectively.  The critical theory on $P_1O_1$ is also modified as a consequence of the $g_{\phi_{int}}$-term. Based on the numerical solution near this critical line, the scaling dimensions of the two competing interaction terms, the $g_{\phi_{int}}$-term and the $g_{\theta_-}$-term both stabilize at 1, indicating the criticality belongs to the Ising universality class. In contrast, the critical behavior on $P_3O_1$ for the $\theta_-$-channel inherits from the critical line $P_3O$ in Fig.~\ref{fig:phase1} since $g_{\phi_{int}}$ flows to 0 and this coupling term is non-consequential.

When close to $O_1O_2$, the energy scales in the even and odd channels are close, hence, the RG processes cannot be decomposed into fast and slow steps any more. 
Since the $g_{\phi_{int}}$-term is the most relevant, it grows 
quickly and overwhelms other terms under sufficiently long RG processes. 
Once $g_{\phi_{int}}$ is renormalized to the strong coupling
region, both $\phi_+$ and $\phi_-$ are pinned, thus the system 
enters into normal state. 
Once it is renormalized to zero, the system is in the SC state and the residual $g_{\theta-}$-term will drive the $Z_2$ symmetry breaking. The transitions on the critical lines across the tri-critical points $O_1$ and $O_2$ are also quite interesting, but we leave the details for future study.

\section{Discussions}
\label{sec:application}
We briefly discuss the application of our theory to the FeTe$_{1-x}$Se$_x$ superconductor, in which evidence to spontaneously time-reversal-symmetry breaking states has been observed by using the high-resolution laser-based 
photo-emission method both in the superconducting and the normal states \cite{zaki2019}.

Following Ref. [\onlinecite{hu2019b}], we consider two superconducting gap 
functions $\Delta_1$ and $\Delta_2$, which possess different pairing 
symmetries and each of them maintains time-reversal symmetry.  It has been argued that the pairing symmetries are constrained to be among $A_{1g(u)}\pm iA_{2g(u)}$, $B_{1g(u)} \pm iB_{2g(u)}$, or $E_{g(u)} \pm iE_{g(u)}$, based on the effects of TRS-breaking pairing on the surface Dirac cone. Here $A,B,E$ denote discrete angular momenta analogous to the $s,d,p$-wave in the continuous case. $g$ and $u$ denote even and odd parities respectively. $A_{1,2}$ means even or odd under vertical plane reflection.
The Ginzburg-Landau free energy is given by,
\begin{equation}
\begin{split}
 \mathcal{F} &=  \alpha_1 \vert \Delta_1\vert^2 + \beta_1\vert \Delta_1\vert^4 + \alpha_2\vert\Delta_2\vert^2 + \beta_2\vert \Delta_2\vert^4 \\
 &+ \kappa\vert \Delta_1\vert^2\vert\Delta_2\vert^2 + 
 \lambda \left( (\Delta_1^\ast\Delta_2)^2 + c.c. \right),
 \end{split}
\end{equation}
where $\alpha_1\approx\alpha_2$ is assumed so that the two pairing 
channels are nearly degenerate as discussed before.
And we focus on the case of $\lambda>0$, where the relative phase between $\Delta_1$ and $\Delta_2$ as $\theta_-=\pm \frac{\pi}{2}$.
Hence, the complex gap function $\Delta_1\pm i\Delta_2$ spontaneously breaks 
time-reversal symmetry.

Since the FeSe$_{1-x}$Te$_x$ superconductor has strong atomic spin-orbit
coupling, as allowed by symmetry, the complex gap function can directly couple to 
the spin magnetization $m_z$ via a cubic coupling term as,
\begin{align}
 \mathcal{F}_M = \alpha_m \vert m_z\vert^2 + i\gamma m_z (\Delta_1\Delta_2^\ast - \Delta_1^\ast\Delta_2),
\end{align}
where $\alpha_m>0$ and $\gamma$ is proportional to the spin-orbit coupling strength \cite{hu2019b}.
This term satisfies both the $U(1)$ symmetry and time-reversal symmetry.
Because of $\alpha_m>0$, the spin magnetization can only be induced by 
the complex gap function via $m_z=\frac{\gamma}{\alpha_m}\vert \Delta_1^\ast\Delta_2\vert\sin\theta_-$ when $\theta_-=\pm \frac{\pi}{2}$.
The development of $m_z$ will gap out the surface Dirac cone as observed in 
the experiment \cite{zaki2019}.
As detailed in Ref.[\onlinecite{hu2019b}], this spontaneous breaking
of TR symmetry can impose a strong constraint on the gap function
symmetry in the FeSe$_{1-x}$Te$_x$ system.  

Furthermore, recent experiment \cite{zaki2019} also shows that the spin-magnetization develops nonzero values even at $T>T_c$,
indicating that TRS breaking already occurs above $T_c$.
It can be understood from the analysis in the main text, where we propose the $Z_2$-breaking normal state.
There are no long-range superconducting orderings, i.e., the $\langle \Delta_1\rangle =\langle \Delta_2\rangle=0 $.
However, the expectation value of the 4-fermion order parameter is nonzero $\langle \Delta_1^\ast\Delta_2\rangle\neq0$ due to the pinning of $\theta_-
=\pm\frac{\pi}{2}$.

\section{Conclusions}
\label{sec:conclusion}

To summarize, we have analyzed the possible symmetry-breaking phases in
the phase fluctuation regime in a two-gap superconductors in 2D. The system has an overall $Z_2\times U(1)$ symmetry, where the $Z_2$ in the $\theta_-$ channel is due to the second order Josephson coupling between the two gaps and the $\theta_+$ channel still has $U(1)$ symmetry. If only the $Z_2$ is broken, then we have the $Z_2$-breaking normal state, which can be either the phase fluctuation induced TRS breaking normal
state or the nematic state, depending on whether the relative phase $\theta_-$
is locked at $\pm\frac{\pi}{2}$, or, at $0$ or $\pi$. On the other hand, if only the $U(1)$ symmetry is broken, then it corresponds to the ordering of the total phase $\theta_+$, even though the two gaps are not individually ordered. This is the quartetting phase, or the so-called $4e$ phase. 

Extensive RG analysis is done by including the more relevant symmetry allowed couplings. 
Not only have we obtained all the four possible phases, including the two interesting intermediate phases in the phase fluctuation regime, 
we also find a direct transition from the $Z_2$-breaking SC state
to the normal state. 
This is because the coupling between half-vortices in the even and odd channels
favors the simultaneous ordering/disordering of the two channels. 

On the experimental side, the TRS-breaking normal phase has been experimentally observed recently in hole-doped $\text{Ba}_{1-x}\text{K}_x\text{Fe}_2\text{As}_2$ \cite{vadim2021state,vadim2022calorimetric}. Furthermore, experimental evidence of the elusive charge-4e state has also been found recently in kagome superconductor $\text{CsV}_3\text{Sb}_5$ \cite{charge4e2022discovery}. The theory presented in this work is based on general symmetry principles. We believe the fluctuation effects and the physical consequences discussed here are quite generic and likely play a role in a wide range of multi-gap superconductors with dominant second-order Josephson couplings. 

{\it Note added:}
Upon the completion of the first version of this manuscript, we became aware of 
two manuscripts on related topics 
Ref.[\onlinecite{fernandes2021charge4e}] and Ref.[\onlinecite{jian2021charge4e}]. Very recently, similar physics have also been discussed in Ref.[\onlinecite{fanyang2023charge}].

\section*{Acknowledgment}
We thank Fan Yang, Yu-Bo Liu and Jing Zhou for helpful discussions. M. Z., H.Y. H. and Y. Z. Y. are supported by a startup funding of UCSD and the National Science Foundation Grant No. DMR-2238360. C.W. is supported by the National Natural Science Foundation of China under the Grants No. 12234016 and No. 12174317. This work has been supported by the New Cornerstone Science Foundation.

\appendix 

\begin{widetext}

\section{The 2D classical $XY$-model and its dual to the sine-Gordon model}
\label{append:duality}
In this section, we review the duality transformation from the $XY$-model to the sine-Gordon model.
We follow Ref. [\onlinecite{herbut2007modern}] to review the duality between the 
$XY$-model and the sine-Gordon model. 
The Hamiltonian of a single-component $XY$-model with the coupling 
constant $J$ is given by,
\begin{align}\label{sm-eq-xy-ham}
 H_{XY} = -J\sum_{\langle i,j\rangle}\cos(\theta_i-\theta_j).
\end{align}
To map the $XY$-model to the sine-Gordon model, we 
start with the Villain approximation,
\begin{equation}\label{sm-eq-vallain}
e^{-K(1-\cos\theta)}\approx \sum_{n=-\infty}^{\infty}e^{-\frac{K}{2}(\theta-2n\pi)^2},
\end{equation}
which is valid when $K$ is large. 
In this case, the dominant contribution comes from the regime
that $\cos\theta\approx 1$, i.e. $\theta \approx 2n\pi$.
Performing Taylor expansion around each of these values, we have $ e^{-K(1-\cos\theta)}\approx \sum_n e^{-\frac{K}{2}(\theta-2n\pi)^2}$.

Using the Villain approximation, the Partition function of the $XY$-model in Eq.~\eqref{sm-eq-xy-ham} is given by
\begin{align}
Z_{XY} =  \int_0^{2\pi} \prod_i\frac{d\theta_i}{2\pi} e^{-\beta H_{XY}}
=\int_0^{2\pi} \prod_i\frac{d\theta_i}{2\pi} e^{\beta J \sum_{\langle i,j\rangle}\cos(\theta_i-\theta_j)}
=\int_0^{2\pi} \prod_i\frac{d\theta_i}{2\pi} \prod_{\langle i,j\rangle}\sum_{m_{ij}}e^{-K/2(\theta_i-\theta_j-2m_{ij}\pi)^2},
\end{align}
where $K=\beta J=J/T$ and the Boltzmann constant is set to be 1 
for simplicity; $m_{ij}$ are integers defined on each link of the 2D lattice.
Now we perform the Hubbard-Stratonovich transformation by introducing the continuous variables $x_{ij}$ defined on each link of the lattice.
The Partition function becomes,
\begin{align}
Z_{XY}=\int_0^{2\pi} \prod_i\frac{d\theta_i}{2\pi} \int_{-\infty}^{\infty}\prod_{<ij>}\sqrt{\frac{2K}{\pi}}dx_{ij}\prod_{\langle i,j\rangle}\sum_{m_{ij}}e^{-\frac{1}{2K}x_{ij}^2-ix_{ij}(\theta_i-\theta_j-2m_{ij}\pi)}.
\end{align}
With the help of the Poisson resummation formula,
\begin{align}
\sum_n\delta(x-nT)=\sum_m \frac{1}{T}e^{i\frac{2m\pi}{T}x},
\end{align}
where $n$ is an integer, the partition function $Z_{XY}$ becomes,
\begin{align}
Z_{XY}&=\int_0^{2\pi} \prod_i\frac{d\theta_i}{2\pi} \int_{-\infty}^{\infty}\prod_{<ij>}\sqrt{\frac{2K}{\pi}}dx_{ij}\prod_{\langle i,j\rangle}e^{-\frac{1}{2K}x_{ij}^2-ix_{ij}(\theta_i-\theta_j)}\sum_n\delta(x_{ij}-n), \\
&\sim\int_0^{2\pi} \prod_i\frac{d\theta_i}{2\pi} \sum_{\{m_{ij}\}}\prod_{\langle i,j\rangle}e^{-\frac{1}{2K}m_{ij}^2-im_{ij}(\theta_i-\theta_j)}.
\end{align}
To perform the above integrals, each $\theta_i$ is extracted from its neighbors,
\begin{align}
Z_{XY}\sim\int_0^{2\pi} \prod_i\frac{d\theta_i}{2\pi} \sum_{\{m_{ij}\}}e^{-\frac{1}{2K}\sum_{i,\hat{\mu}}m_{i,\hat{\mu}}^2-i
	\sum_{i,\hat{\mu}}(m_{i,\hat{\mu}}-m_{i,-\hat{\mu}})\theta_i},
\end{align}
where $\hat{\mu}=\hat x, \hat y$ denotes the lattice unit vectors along
the bond directions. 
Now the angles $\theta_i$ can be integrated out,
\begin{align}
Z_{XY}\sim \sum_{\{m_{ij}\}}e^{-\frac{1}{2K}
\sum_{i,\hat{\mu}}m_{i,\hat{\mu}}^2} 
\prod_i\delta \left(\sum_{\hat{\mu}} (m_{i,\hat{\mu}}-m_{i,-\hat{\mu}})\right),
\end{align}
where the $\delta$-function here is the the Kronecker $\delta$. 

Each integer $m_{ij}$ defined on the link can be treated as a current flown into and out of the connected lattice sites, and the $\delta$-function here basically says the current through each site is conserved. This conservation constraint is naturally satisfied if we define another set of integers $\{n_i\}$ at the sites of the dual lattice, i.e. the centers of the plaquettes of the original lattice,
\begin{equation}
\begin{split}
&m_{i,\hat{x}}=n_{i+\hat{x}+\hat{y}}-n_{i+\hat{x}},\\
&m_{i,\hat{y}}=n_{i+\hat{y}}-n_{i+\hat{x}+\hat{y}},\\
&m_{i-\hat{x},\hat{x}}=n_{i+\hat{y}}-n_{i},\\
&m_{i-\hat{y},\hat{y}}=n_{i}-n_{i+\hat{x}}.
\end{split}
\end{equation}
With the new set of integers, the partition function now becomes,
\begin{align}
Z_{XY}\sim \sum_{\{n_i\}}e^{-\frac{1}{2K}\sum_{i,\hat{\mu}}(n_{i+\hat{\mu}}-n_i)^2}.
\end{align}
Comparing with the original partition function, we notice that the temperature has been inverted because $K\to 1/K$, and continuous variables has been replaced by integer variables. However, we can use Poisson summation to go back to continuous variables. Therefore,
\begin{align}
Z_{XY}\sim \int \prod_i d\phi_i \sum_{\{n_i\}}e^{-\frac{1}{2K}\sum_{i,\hat{\mu}}(\phi_{i,\hat{\mu}}-\phi_i)^2}\prod_i\delta(\phi_i-n_i)
=\int \prod_i d\phi_i \sum_{\{n_i\}}e^{-\frac{1}{2K}\sum_{i,\hat{\mu}}(\phi_{i,\hat{\mu}}-\phi_i)^2-i2\pi\sum_in_i\phi_i},
\end{align}
After adding the chemical potential term, the Partition function becomes,
\begin{align}
Z_{XY}\sim\int \prod_i d\phi_i \sum_{\{n_i\}}e^{-\frac{1}{2K}\sum_{i,\hat{\mu}}(\phi_{i,\hat{\mu}}-\phi_i)^2-i2\pi\sum_in_i\phi_i+\mathrm{ln}y\sum_in_i^2}.
\end{align}
Next we perform the summation over $\{n_i\}$ by using the following identity,
\begin{eqnarray}
\sum_{\{n_i\}}e^{-i2\pi\sum_in_i\phi_i+\mathrm{ln}y\sum_in_i^2}
&=&\prod_i\sum_{n_i=0,\pm 1,...}y^{n_i^2}e^{-i2\pi n_i\phi_i}
=\prod_i(1+2y\cos2\pi\phi_i+O(y^2))\nonumber \\
=e^{2y\sum_i\cos2\pi \phi_i}.
\end{eqnarray}
The partition function eventually becomes the form of the sine-Gordon model,
\begin{align}
Z_{XY}\sim \int \prod_i d\phi_i e^{-\frac{1}{2K}\sum_{i,\hat{\mu}}(\phi_{i,\hat{\mu}}-\phi_i)^2+2y\sum_icos2\pi \phi_i}.
\end{align}

\section{RG equations from operator product expanions}
\label{append:RG-derivation}
\subsection{Scaling dimensions}
\label{append:scaling dimension}
In this part we use the operator product expansion (OPE) to calculate the scaling dimensions of the coupling terms consisting of vertex operators of the form $\cos\beta\phi$ in the free bosonic field $\phi$ and the vertex operators $\cos\beta\theta$ in the dual field $\theta$, based on the free Lagrangian $\mathcal{L}_0=\frac{1}{2K}(\partial_{\mu}\phi)^2$. Notice that the Luttinger parameter $K$ in the results presented below have to be accordingly scaled in order to be used for the theory in Eq.~(\ref{Eq:lagrangian2}). 

We start with the correlation functions of the following vertex operators.
Following the notation in Ref. [\onlinecite{shankar2017quantum}], the 
correlation function is given by,
\begin{align}
G_{\beta}(x-y)\equiv\langle \mathrm{e}^{i\beta \phi(x)}\mathrm{e}^{-i\beta \phi(y)} \rangle.
\end{align}
By using the operator identity: $\mathrm{e}^A\mathrm{e}^B:=:\mathrm{e}^{A+B}:\mathrm{e}^{\langle AB+\frac{A^2+B^2}{2} \rangle}$, where $:\hat{O}:$ means normal ordering, we have
\begin{equation}
\begin{split}
G_{\beta}(x-y)&=\langle :\mathrm{e}^{i\beta(\phi(x)-\phi(y))}: \rangle \mathrm{e}^{-\frac{\beta^2}{2}\langle (\phi(x)-\phi(y))^2\rangle}
=\mathrm{e}^{\beta\langle \phi(x)\phi(y)-\phi^2(x)\rangle}
=\lim_{l \to 0}\left( \frac{l^2}{l^2+(x-y)^2}\right)^{\frac{\beta^2K}{4\pi}},
\end{split}
\end{equation}
where $l$ here is the short distance cutoff.
The following fact is used to derive the above equation,
\begin{align}
\langle \phi(x)\phi(y)-\phi^2(x)\rangle=-\frac{K}{2\pi}\mathrm{ln}\frac{l^2}{l^2+(x-y)^2}.
\end{align}
Similarly, we have for the dual field $\theta$:
\begin{align}
\langle \theta(x)\theta(y)-\theta^2(x)\rangle=-\frac{1}{2\pi K}\mathrm{ln}\frac{l^2}{l^2+(x-y)^2}.
\end{align}
Therefore, we are able to obtain the following correlation functions for 
two different types of vertex operators:
\begin{equation}
\begin{split}
&\langle \mathrm{e}^{i\beta \phi(x)}\mathrm{e}^{-i\beta \phi(y)} \rangle\sim |x-y|^{-\frac{\beta^2K}{2\pi}},\\
&\langle \mathrm{e}^{i\beta \theta(x)}\mathrm{e}^{-i\beta \theta(y)} \rangle\sim |x-y|^{-\frac{\beta^2}{2\pi K}},
\end{split}
\end{equation}
based on which the scaling dimensions of the vertex operators can 
be calculated.

By taking
$\cos\beta\phi=\frac{1}{2}(\mathrm{e}^{i\beta\phi}+\mathrm{e}^{-i\beta\phi})$, then
\begin{equation}
\begin{split}
\langle \mathrm{cos}\beta \phi(x)\mathrm{cos}\beta\phi(y)\rangle
&=\frac{1}{4}\left(\langle \mathrm{e}^{i\beta\phi(x)}\mathrm{e}^{i\beta\phi(y)} \rangle+\langle \mathrm{e}^{i\beta\phi(x)}\mathrm{e}^{-i\beta\phi(y)} \rangle+\langle \mathrm{e}^{-i\beta\phi(x)}\mathrm{e}^{i\beta\phi(y)} \rangle+\langle \mathrm{e}^{-i\beta\phi(x)}\mathrm{e}^{-i\beta\phi(y)} \rangle \right)
\nonumber \\
&\sim |x-y|^{-\frac{\beta^2K}{2\pi}},
\end{split}
\end{equation}
where we have used the fact that $\langle \mathrm{e}^{i\beta_1\phi(x_1)}...\mathrm{e}^{i\beta_N\phi(x_N)}\rangle=0$ in the thermodynamic limit when $\sum_{n=1}^N\beta_n\neq 0$ [\onlinecite{tsvelik2007quantum}]. From this we conclude that the scaling dimension of the $\cos\beta\phi$ term is $\frac{\beta^2K}{4\pi}$. Similarly the $\cos\beta\theta$ term has scaling dimension $\frac{\beta^2}{4\pi K}$. Using these results, the composite operators consisting of this two types of basic vertex operators, like the ones in the main text, can be readily calculated.

\subsection{The one-loop correction}
For the one-loop corrections for the RG equations, we consider first the 
simple case where the free bosonic Lagrangian $\mathcal{L}_0=\frac{1}{2K}(\partial_{\mu}\phi)^2$ is perturbed 
by a generic vortex term $\mathcal{L}'=\frac{g_{\phi}}{l^{D-\Delta_{\phi}}}\mathrm{cos}\beta \phi+\frac{g_{\theta}}{l^{D-\Delta_{\theta}}}\mathrm{cos}\alpha\theta$, where the short-distance cutoff $l$ is restored to make the couplings dimensionless or scale invariant [\onlinecite{fradkin2013field}].
The partition function can then be expanded as the following:
\begin{equation}\label{sm-eq-partition-phi-theta}
\begin{split}
Z=\int D[\phi]e^{-S}
&=Z^*\Big(1+\int dx\frac{g_{\phi}}{l^{D-\Delta_{\phi}}}\langle\mathrm{cos}\beta\phi\rangle+\int dx\frac{g_{\theta}}{l^{D-\Delta_{\theta}}}\langle\mathrm{cos}\alpha\theta\rangle
+\frac{1}{2}\int dxdy\frac{g_{\phi}g_{\theta}}{l^{2D-\Delta_{\phi}-\Delta_{\theta}}}\langle\mathrm{cos}\beta\phi(x)\mathrm{cos}\alpha\theta(y)\rangle\\
&+\frac{1}{2}\int dxdy\frac{g_{\phi}^2}{l^{2D-2\Delta_{\phi}}}\langle\mathrm{cos}\beta\phi(x)\mathrm{cos}\beta\phi(y)\rangle
+\frac{1}{2}\int dxdy\frac{g_{\theta}^2}{l^{2D-2\Delta_{\theta}}}\langle\mathrm{cos}\alpha\theta(x)\mathrm{cos}\alpha\theta(y)\rangle+O(g^3)\Big),
\end{split}
\end{equation}
where $Z^*$ represents the free theory partition function. 
As we know, the conformal invariance of the free theory requires that the cross term corresponding to $g_{\phi}g_{\theta}$ vanishes at the one-loop level because the $g_{\phi}$ and the $g_{\theta}$ terms in general have different scaling dimensions.
So we only need to consider the $g_\phi^2$ and $g_\theta^2$ terms.

Firstly, consider the $g_\phi^2$ term. The OPE in terms of $e^{i\beta\phi}$ is given by [\onlinecite{fradkin2013field}],
\begin{equation}
:\mathrm{e}^{i\beta \phi(x)}::\mathrm{e}^{-i\beta \phi(y)}:=\frac{1}{|x-y|^{2\Delta_{\phi}}}-\frac{1}{|x-y|^{2\Delta_{\phi}-2}}\frac{\beta^2}{2}:(\partial_\mu\phi)^2:,
\end{equation}
\begin{equation}
:\mathrm{e}^{\pm i\beta \phi(x)}::\mathrm{e}^{\pm i\beta \phi(y)}:=\frac{1}{|x-y|^{-2\Delta_{\phi}}} :\mathrm{e}^{\pm i2\beta\phi(x)}:,
\end{equation}

\begin{equation}
    :\mathrm{e}^{i\alpha \theta(x)}::\mathrm{e}^{-i\alpha \theta(y)}:=\frac{1}{|x-y|^{2\Delta_{\theta}}}-\frac{1}{|x-y|^{2\Delta_{\theta}-2}}\frac{\alpha^2}{2}:(\partial_\mu\theta)^2:,
\end{equation}
\begin{equation}
    :\mathrm{e}^{\pm i\alpha \theta(x)}::\mathrm{e}^{\pm i\alpha \theta(y)}:=\frac{1}{|x-y|^{2\Delta_{\theta}}}:\mathrm{e}^{\pm i2\alpha\theta(x)}:,
\end{equation}
where it is understood that $|x-y|\to 0$. 
Therefore,
\begin{equation}
\begin{split}
:\mathrm{cos}\beta\phi(x)::\mathrm{cos}\beta\phi(y):&=\frac{1}{4}:(\mathrm{e}^{i\beta \phi(x)}+\mathrm{e}^{-i\beta \phi(x)})::(\mathrm{e}^{i\beta \phi(y)}+\mathrm{e}^{-i\beta \phi(y)}):\\
&=\frac{1/2}{|x-y|^{2\Delta_{\phi}}}-\frac{1/2}{|x-y|^{2\Delta_{\phi}-2}}\frac{\beta^2}{2}:(\partial_\mu\phi)^2:+\frac{1/2}{|x-y|^{-2\Delta_{\phi}}}:\cos2\beta\phi(x):,
\end{split}
\end{equation}
and similarly, 
\begin{equation}
:\mathrm{cos}\alpha\theta(x)::\mathrm{cos}\alpha\theta(y):=\frac{1/2}{|x-y|^{2\Delta_{\theta}}}-\frac{1/2}{|x-y|^{2\Delta_{\theta}-2}}\frac{\alpha^2}{2}:(\partial_\mu\theta)^2:+\frac{1/2}{|x-y|^{-2\Delta_{\theta}}}:\cos2\alpha\theta(x):.
\end{equation}
For the $g_{\phi}^2$ term in Eq.~\eqref{sm-eq-partition-phi-theta},
$\frac{1}{2}\int dxdy\frac{g_{\phi}^2}{l^{2D-2\Delta_{\phi}}}\langle\mathrm{cos}\beta\phi(x)\mathrm{cos}\beta\phi(y)\rangle$, which gives rise to the one-loop correction to the $:(\partial_{\mu}\phi)^2:$ term, becomes
\begin{equation}
-\frac{\beta^2}{8}\int dxdy\frac{g_{\phi}^2}{l^{2D-2\Delta_{\phi}}}|x-y|^{-2\Delta_\phi+2} \langle :(\partial_{\mu}\phi)^2:\rangle
=-\frac{\beta^2}{8}\int dx\frac{g_{\phi}^2}{l^{2D-2\Delta_{\phi}}}\langle :(\partial_{\mu}\phi)^2:\rangle\int dy|x-y|^{-2\Delta_\phi+2}.
\end{equation}
Now we do a change of scale by changing the cutoff $l\to l+\delta l=(1+\delta \ln l)l$. This means the domain of the above integration is changed from $|x-y|>l$ to $|x-y|>(1+\delta \ln l)l$. Therefore, the corresponding change in the above integration becomes,
\begin{equation}
\frac{\beta^2}{8}\int dx\frac{g_{\phi}^2}{l^{2D-2\Delta_{\phi}}}\langle :(\partial_{\mu}\phi)^2:\rangle\int_{l<|x-y|<(1+\delta \ln l)l} dy|x-y|^{-2\Delta_\phi+2},
\end{equation}
which in the case of $D=2$ is,
\begin{align}
\frac{\beta^2\pi}{4}g_{\phi}^2\delta \ln l \int dx\langle :(\partial_{\mu}\phi)^2:\rangle.
\end{align}
Comparing with the kinetic term $\frac{1}{2K}\int dx (\partial_{\mu}\phi)^2$, we obtain the correction of $K$ due to the $g_{\phi}$ term,
\begin{equation}
\frac{d(1/K)}{d\ln l}=\frac{\pi \beta^2}{2}g_{\phi}^2\Rightarrow \frac{d K}{d\ln l}=-\frac{\pi \beta^2K^2}{2}g_{\phi}^2.
\end{equation}
The contribution from the $g_{\theta}\cos\alpha\theta$ term can be similarly obtained as, 
\begin{align}
 \frac{dK}{d \ln l}=\frac{\pi \alpha^2}{2}g_{\theta}^2.
\end{align}

\subsection{Derivation of the RG equations}
Using the basic ingredients above, we can proceed to work out the full RG equations presented in the main text. For the free theory given by,
\begin{equation}
\mathcal{L}_0=\frac{1}{4K_+}(\partial_{\mu}\phi_+)^2+\frac{1}{K_-}
(\partial_{\mu}\phi_-)^2\equiv \frac{1}{2\tilde{K}_+}(\partial_{\mu}\phi_+)^2+\frac{1}{2\tilde{K}_-}
(\partial_{\mu}\phi_-)^2,
\end{equation}
where we have redefined the Luttinger parameters $\tilde{K}_+\equiv 2K_+,\tilde{K}_-\equiv K_-/2$, so that the Lagrangian takes the standard normalization convention and the results derived from the previous section can be directly carried over. 
We have the following scaling dimensions for the different interaction terms:
\begin{itemize}
    \item For $\cos\beta\phi_+$: $\Delta_{\phi_+}=\frac{\beta^2\tilde{K}_+}{4\pi}$;
     \item For $\cos\beta\phi_-$: $\Delta_{\phi_-}=\frac{\beta^2\tilde{K}_-}{4\pi}$;
    \item For $\cos\alpha\theta_-$: $\Delta_{\theta_-}=\frac{\alpha^2}{4\pi \tilde{K}_-}$;
    \item For $\cos\beta\phi_+\cos\alpha\phi_-$: $\Delta_{\phi_+\phi_-}=\frac{1}{4\pi}\left(\beta^2\tilde{K}_++\alpha^2\tilde{K}_-\right)$.
\end{itemize}
The scaling dimensions above give us the tree-level flow equations. For the loop-level correction of the Luttinger parameters, we again make use of the OPEs. The OPE
\begin{equation}
:\mathrm{cos}\beta\phi_+(x)::\mathrm{cos}\beta\phi_+(y):
=\frac{1/2}{|x-y|^{2\Delta_{\phi_+}}}-\frac{1/2}{|x-y|^{2\Delta_{\phi_+}-2}}\frac{\beta^2}{2}:(\partial_\mu\phi_+)^2:,
\end{equation}
gives the following correction after repeating the real-space renormalization,
\begin{equation}
    \delta\left(\frac{1}{2\tilde{K}_+}\right)=\frac{1}{2}\cdot \frac{\beta^2}{4}g_{\phi_+}^2\cdot 2\pi \delta\left(\ln l\right)\Rightarrow \delta \tilde{K}_+=-\frac{\pi}{2}\beta^2\tilde{K}_+^2g_{\phi_+}^2\delta\left(\ln l\right).
\end{equation}
The OPE
\begin{equation}
:\mathrm{cos}\beta\phi_-(x)::\mathrm{cos}\beta\phi_-(y):
=\frac{1/2}{|x-y|^{2\Delta_{\phi_-}}}-\frac{1/2}{|x-y|^{2\Delta_{\phi_-}-2}}\frac{\beta^2}{2}:(\partial_\mu\phi_-)^2:,
\end{equation}
gives the following correction to $\tilde{K}_-$,
\begin{equation}
    \delta\left(\frac{1}{2\tilde{K}_-}\right)=\frac{1}{2}\cdot \frac{\beta^2}{4}g_{\phi_-}^2\cdot 2\pi \delta\left(\ln l\right)\Rightarrow \delta \tilde{K}_-=-\frac{\pi}{2}\beta^2\tilde{K}_-^2g_{\phi_-}^2\delta\left(\ln l\right).
\end{equation}
The OPE 
\begin{equation}
:\mathrm{cos}\alpha\theta_-(x)::\mathrm{cos}\alpha\theta_-(y):=\frac{1/2}{|x-y|^{2\Delta_{\theta_-}}}-\frac{1/2}{|x-y|^{2\Delta_{\theta_-}-2}}\frac{\alpha^2}{2}:(\partial_\mu\theta_-)^2:,
\end{equation}
gives the following correction to $K_-$,
\begin{equation}
    \delta\left(\frac{\tilde{K}_-}{2}\right)=\frac{1}{2}\cdot \frac{\alpha^2}{4}g_{\theta_-}^2\cdot 2\pi \delta\left(\ln l\right)\Rightarrow \delta \tilde{K}_-=\frac{\pi\alpha^2}{2}g_{\theta_-}^2\delta\left(\ln l\right).
\end{equation}
The OPE
\begin{equation}
\begin{split}
    &:\cos\beta\phi_+(x)\cos\alpha\phi_-(x)::\cos\beta\phi_+(y)\cos\alpha\phi_-(y):\\
    &=\frac{1}{16}\sum_{\eta_{1,2,3,4}=\pm 1}:\mathrm{e}^{i\left(\eta_1\beta\phi_+(x)+\eta_2\alpha\phi_-(x)\right)}::\mathrm{e}^{i\left(\eta_3\beta\phi_+(y)+\eta_4\alpha\phi_-(y)\right)}:\\
    &\Longrightarrow -\frac{1/8}{|x-y|^{2\Delta_{\phi_+}+2\Delta_{\phi_-}-2}}\left(\beta^2:(\partial_\mu\phi_+)^2:+\alpha^2:(\partial_\mu\phi_+)^2:\right)\\
    &\quad \quad +\frac{1/4}{|x-y|^{-2\Delta_{\phi_+}+2\Delta_{\phi_-}}}:\cos2\beta\phi_+:+\frac{1/4}{|x-y|^{2\Delta_{\phi_+}-2\Delta_{\phi_-}}}:\cos2\alpha\phi_-:,
\end{split}
\end{equation}
which generates the following renormalizations,
\begin{equation}
    \begin{split}
       & \delta\left(\frac{1}{2\tilde{K}_+}\right)=\frac{1}{2}\cdot \frac{\beta^2}{8}g_{\phi_{int}}^2\cdot 2\pi \delta\left(\ln l\right)\Rightarrow \delta \tilde{K}_+=-\frac{\pi\beta^2\tilde{K}_+^2}{4}g_{\phi_{int}}^2\delta\left(\ln l\right),\\
       &\delta\left(\frac{1}{2\tilde{K}_-}\right)=\frac{1}{2}\cdot \frac{\alpha^2}{8}g_{\phi_{int}}^2\cdot 2\pi \delta\left(\ln l\right)\Rightarrow \delta \tilde{K}_-=-\frac{\pi\alpha^2\tilde{K}_-^2}{4}g_{\phi_{int}}^2\delta\left(\ln l\right),\\
       &\delta g_{\phi_\pm}=\frac{1}{2}\cdot \frac{1}{4}g_{\phi_{int}}^2\cdot 2\pi \delta\left(\ln l\right)=\frac{\pi}{4}g_{\phi_{int}}^2\delta\left(\ln l\right).
    \end{split}
\end{equation}

Combining the contributions from the different interaction terms, we eventually arrive at the RG equations presented in the main text.

\section{K-matrix formulation of Luttinger liquid}
\label{append::Kmatrix}
In this section, we review the K-matrix formulation of the Luttinger liquid. In this framework, a Luttinger liquid is treated as the boundary of a higher-dimensional bulk and the K-matrix contains topological information about the bulk. In particular, using the K-matrix it is straightforward to calculate the braiding statistics between the various vertex operators that represent the charges, vortices or their combinations. This is a useful way to rule out non-local operators when writing down the Lagrangian based on symmetry considerations.
\subsection{One pair of boson and dual boson}
To warm up for the case of two coupled Luttinger liquids in our paper, we look at the simpler case of one Luttinger liquid consisting of the boson field $\theta$ and its dual $\phi$. By defining $\Phi\equiv (\theta,\phi)^\text{T}$, the free Lagrangian density is given by,
\begin{equation}
    \mathcal{L}_0=\frac{1}{4\pi}\left(\partial_t \Phi^\text{T} \text{K} \partial_x\Phi+\partial_x \Phi^\text{T} \text{V} \partial_x\Phi\right),
    \label{eq::Lagrangian-Kmatrix}
\end{equation}
where the K here is not to be confused with the Luttinger parameter $K$ that appears in the rest part of the paper. The K-matrix is given by $\text{K}=\sigma^1$ and the V-matrix is given by $\text{V}=\sigma^0$, where the $\sigma^\mu$ with $\mu=0,1,2,3$ are the Pauli matrices. In canonical quantization, the conjugate momentum of the $\theta$ field is given by,
\begin{equation}
    \Pi=\frac{\delta\mathcal{L}_0}{\delta\partial_t\theta}=\frac{1}{2\pi}\partial_x\phi,
\end{equation}
with the canonical commutation given by $[\theta(t,x),\Pi(t,y)]=i\delta(x-y)$, or equivalently, $[\theta(t,x),\partial_y\phi(t,y)]=2\pi i\delta(x-y)$. 

We have two basic types of vertex operators $\mathrm{e}^{i\theta}$ and $\mathrm{e}^{i\phi}$, whose charge vectors are given by $l_\theta=(1,0)^\text{T}$ and $l_\phi=(0,1)^\text{T}$ respectively. Then the braiding statistics between the two vertex operators is given by,
\begin{equation}
    2\pi l_\theta^\text{T}\text{K}^{-1}l_\phi=2\pi,
\end{equation}
which simply states the fact that if we move a charge around its vortex, then it picks up a phase of $2\pi$. Here we take $\mathrm{e}^{i\theta}$ to be the charge operator and the $\mathrm{e}^{i\phi}$ to be the vortex operator to be consistent with the notation of the main text. Notice however, that in the normalization convention of the main text, the vortex is given by $\mathrm{e}^{i2\pi\phi}$ instead, so there is a factor of $2\pi$ in the field rescaling for $\phi$. In the convention used here, $\theta$ and $\phi$ are put on equal footing, both without the $\pi$ factors. The normalization convention does not change the essential physics we discuss. 
\subsection{Two coupled Luttinger liquids}
Now we move on to two coupled Luttinger liquids, which would correspond to two coupled XY-models. Choosing the basis $\Phi=(\theta_1,\phi_1,\theta_2,\phi_2)^\text{T}$, the Lagrangian density takes the same form as in Eq~(\ref{eq::Lagrangian-Kmatrix}), but the new K-matrix and V-matrix are given by,
\begin{equation}
    \text{K}=\begin{pmatrix}\sigma^x & 0\\0& \sigma^x\end{pmatrix},\quad \text{V}=\begin{pmatrix}\sigma^0 & 0\\0& \sigma^0\end{pmatrix}.
\end{equation}
Then we have the following charge vectors,
\begin{equation}
    l_{\theta_1}=(1,0,0,0)^\text{T},\quad  l_{\phi_1}=(0,1,0,0)^\text{T},\quad
    l_{\theta_2}=(0,0,1,0)^\text{T},\quad  l_{\phi_2}=(0,0,0,1)^\text{T}.
\end{equation}
Under the basis transformation used in the main text,
\bea
\theta_{+}&\equiv& (\theta_1+\theta_2)/2, 
~\theta_{-}\equiv \theta_1-\theta_2 \nn \\
\phi_{+} &\equiv& \phi_1+\phi_2, \ \ \
\ \ \ \phi_-\equiv (\phi_1-\phi_2)/2,
\eea
the charge vectors for the new fields are given by,
\begin{equation}
    l_{\theta_+}=(\frac{1}{2},0,\frac{1}{2},0)^\text{T},\quad l_{\phi_+}=(0,1,0,1)^\text{T},\quad l_{\theta_-}=(1,0,-1,0)^\text{T},\quad l_{\phi_-}=(0,\frac{1}{2},0,-\frac{1}{2})^\text{T}.
\end{equation}
In a similar fashion, the braiding between the fields and the dual fields are given by,
\begin{equation}
    \begin{split}
        &  2\pi l_{\theta_+}^\text{T}\text{K}^{-1}l_{\phi_+}=2\pi,\quad 2\pi l_{\theta_-}^\text{T}\text{K}^{-1}l_{\phi_-}=2\pi,\\
        & 2\pi l_{\theta_+}^\text{T}\text{K}^{-1}l_{\phi_-}=0,\quad 2\pi l_{\theta_-}^\text{T}\text{K}^{-1}l_{\phi_+}=0,
    \end{split}
\end{equation}
i.e. the braiding between fields from different channels vanishes, as it should be. 

Now we are ready to check the locality of the various terms appearing in the Lagrangian, i.e. whether their braiding with the original local physical fields $\theta_{1,2}$ are integer multiples of $2\pi$. 
\begin{itemize}
    \item $\cos\phi_+$ (Note again that this term is the $\cos2\pi\phi_+$ in the main text): The charge vector is $(0,1,0,1)^\text{T}$, and its braiding with $\theta_{1,2}$ are both $2\pi$. Higher order terms are therefore also allowed.
    \item $\cos2\theta_-$: The charge vector is $(1,0,1,0)^\text{T}$, and its braiding with $\theta_{1,2}$ are both 0.
    \item $\cos\phi_-$ (equivalent to $\cos2\pi \phi_-$ in the main text) : The charge vector is $(0,\frac{1}{2},0,-\frac{1}{2})^\text{T}$, and its braiding with $\theta_{1,2}$ are $\pm \pi$ respectively, i.e. not integer multiple of $2\pi$, hence not allowed. 
    \item $\cos2\phi_-$ (equivalent to $\cos4\pi \phi_-$ in the main text): The charge vector is $(0,1,0,-1)^\text{T}$, and its braiding with $\theta_{1,2}$ are $\pm 2\pi$.
    \item $\cos\frac{1}{2}\phi_+\cos\phi_-$ (equivalent to $\cos\pi\phi_+\cos2\pi\phi_-$ in the main text): The charge vector is given by $(0,1,0,0)^\text{T}$, whose braiding with $\theta_{1,2}$ are $2\pi$ and 0 respectively, hence it is local and allowed, even though neither $\cos\frac{1}{2}\phi_+$ nor $\cos\phi_-$ is allowed separately. This is consistent with the fact that this term comes from the sum of the original two local vortex terms $\cos\phi_1$ and $\cos\phi_2$. 
\end{itemize}
\end{widetext}
\bibliographystyle{apsrev4-1}
\bibliography{bib}

\end{document}